\documentclass[12pt]{article}
\usepackage{graphicx,latexsym,epstopdf}

\begin{document}

\begin{titlepage}

\vspace{3mm}

\begin{center}
{\Large {\bf Birth and Growth \\ of \\ Two-dimensional Universe
}}
\end{center}

\vspace{3mm}

\begin{center}
{\sc Tetsuyuki Yukawa}
\end{center}

\begin{center}
{\it Institute of Particle and Nuclear Studies, \\KEK, Tsukuba, Ibaraki 305-0801, Japan} \\
{\it and}\\
{\it The Graduate University for Advanced Studies (Sokendai), \\Hayama, Kanagawa 240-0193, Japan} \\
\end{center}

\begin{abstract}
A master equation for the evolution of two-dimensional universe is derived based on the simplicial quantum gravity regarding the evolution as the Markov process of a space-time lattice. 
Three typical phases, expanding, elongating and collapsing phase, which have been found in the numerical simulation, are studied together with their boundaries, analytically. 
Asymptotic solutions of the evolution equation for statistical quantities, such as the averaged area, the boundary length, correlations of their fluctuations, are obtained for each phase and boundary.
After introducing a physical time the cosmological significance of each phase is discussed.  

\end{abstract}

\end{titlepage}
\section{Introduction}
\noindent
Recent observations and their analyses on anisotropies of the cosmic microwave background \cite{COBE, WMAP} have revealed the large scale correlation between two points separated far beyond the event horizon.
It is considered as a relic of quantum fluctuation produced during the period when the universe is a quantum mechanical size. 
The full sky map of universe tempts us to imagine that we are glancing at the stage where the universe begins to explode. 
In order to describe this period on the scientific ground it is inevitable to have a tool which is capable of handling the creation of space, time, and matter all at the same time. 

It is quantum gravity which is aimed to describe such a circumstance.
Quantum gravity is also expected to be the theory for quantizing the gravitational field.
However, incompleteness of various theoretical attempts in the 4-dimensional space-time by standard quantization methods made most of the investigations turn toward a totally new direction, {\it i.e.} the super-string theory, except a few investigations \cite{CG,HS}.
On the other hand, as far as the creation of space, time, and matter is concerned, there exist consistent theories of quantum gravity in the 2-dimensional space-time, where no quantization of the gravitational field is necessary.
It is our aim of this article to give an insight on birth and growth of the universe through the 2-dimensional quantum gravity.

There are several standard methods of the 2-dimensional quantum gravity, for examples, the Liouville theory\cite{LIOUVILLE} based on the standard field theory, the matrix model\cite{MATRIX-THEORY} representing the combinatorial geometry\cite{TUTTE}, and the dynamical triangulation\cite{DT2D} as its numerical realization.
Recently, we have proposed a scenario of the initial stage of universe in terms of the simplicial quantum gravity\cite{qc4,HY}. 
With a few fundamental axioms a d-dimensional universe has been constructed as a set of complexes of d-simplices.
Numerical simulations have been performed in $ 2 $, and $ 4 $ dimensions by applying techniques of the dynamical triangulation.
They have indicated that there are basically three types of universes, which have been called the collapsing, the expanding and the elongating phase. 
For the 2-dimensional case numerical results have been compared to predictions of the boundary Liouville theory\cite{FZZ}, and shown to reproduce the expected results.

However, those comparisons have been limited within the collapsing phase, where the stable vacuum exists and the standard field theory technique is applicable. 
In order to analyze non-static phases, in which the spaces keep growing indefinitely, we need to find an appropriate method that can handle a system without the stable vacuum.
In this article we propose a method to treat such cases based on techniques of the non-equilibrium statistical mechanics.
This study shows that the 2-dimensional simplicial quantum gravity enables to describe most of the essential features of the 4-dimensional universe, such as the inflation and the big bang, even under the limited condition of dimensionality.

There exist a few attempts for creation of universe in the similar standpoint as ours, expecting the underlying discrete nature of space-time, such as the causal dynamical triangulation (CDT)\cite{CDT} and the loop quantum gravity (LQG)\cite{LQG}.
When one consider the very beginning of universe, one inevitably give the initial hypothesis which cannot be proven.
The main difference of those models comes from a point where the model starts.
Both the CDT and the LQG assume the existence of the Lorentzian space-time and quantum dynamics before the birth of universe, while we consider all the physical laws appear at or after the birth of universe. 
Thus our future task will be to derive the dynamical law and the Lorentzian invariance of dynamics.

In the next section (Sec.2) the scenario of creation of universe is reviewed within the $2$-dimensional quantum gravity. 
I derive a master equation as the continuous limit of the dynamical triangulation with a boundary. 
In Sec.3, I study the statistical property of static universe through calculating the partition function as a generating function for the average of area, boundary length, and correlation functions of fluctuations.
Sec.4 is devoted to overview possible non-static solutions of the master equation.
In Sec.5, I solve equations of the motion for averaged quantities derived from the master equation for the expanding phase and its boundaries.
Following section (Sec.6) describes the elongating phase, where the boundary length stays constant while the area keeps increasing.
In Sec.7 two special states of universe are discussed.
One is a universe with the minimum boundary length, which may be called the polymer state, appearing for a specific value of the parameter related to the matter central charge, and the other is a boundary state along the line locating between the expanding and elongating phases.
Sec.8 is devoted to discuss evolution of the universe in $ 1+1 $ dimension by introducing a physical time, and show that it is possible to interpret each phase as a state of actual universe at various stages of its evolution, such as the inflation, the big bang, the stationary, and the crunch.
Last section (Sec.9) summarizes my aim, accomplishments, and basic problems left to be solved.
Three appendices are in order for the master equation in the continuous limit and two basic equations of the motion referred in the text. 

\section{Master equation for the evolution of universe}
\setcounter{equation}{0}
\noindent
We define the 2-dimensional universe to be a set of all the possible simplicial complexes of triangles with one boundary.
We shall call a complex as a configuration, or sometimes a world.
It is possible to classify those simplicial manifolds by the number of handles $ g $.
In this article we limit our worlds only to those manifolds of the simple disk topology without handles, for simplicity.
Each complex has its neighbors which can be obtained by attaching or detaching one unit triangle.
In 2-dimension, there are, in general, four directions of neighbors for each configuration, either increasing or decreasing the area and boundary length by one unit.
These procedures are called moves.

In practice, a series of configurations is generated by selecting randomly one of those possible neighbors connected by a move at each step.
We start with any but normally the simplest configuration made of one elementary triangle to create a Markov chain of the length $ N $,
$$
\alpha_1 \rightarrow \alpha_2 \rightarrow \cdots \rightarrow \alpha_N .
$$ 
A configuration is specified by a set of 'quantum' numbers, $ \alpha=$ ($N_2$, $\tilde{N}_1$, $k$), namely, the total number of elementary triangles, $ N_2$, the number of edges on the boundary, $ \tilde{N}_1$, and the quantum number $k$ identifying a specific diagram with a fixed ($ N_2$, $ \tilde{N}_1$). 
A selected move is accepted if the resultant configuration satisfies following manifold conditions:
{\it 1}) at most two triangles attach through one edge, and {\it 2}) triangles sharing one vertex form a 2-disk, or a semi-disk if the vertex is on the boundary.

Mathematically the dynamical triangulation can be described by a master equation for the probability $ p_{\alpha}[n] $ of a configuration $ \alpha $ at the $ n $-th step, which is expressed as
\begin{equation}
p_{\alpha}[n+1] = p_{\alpha}[n] + \sum_{\beta(\alpha)} \{p_{\beta}[n] w_{\beta \rightarrow \alpha}-p_{\alpha}[n] w_{ \alpha \rightarrow \beta} \}, 
\end{equation}
where the sum over $ \beta(\alpha) $ is understood to take all those neighbors $\{ \beta \}$ connected by one of the four moves from a configuration $ \alpha $ .
The transition probability $ w_{ \alpha \rightarrow \beta} $ is chosen so that in the case when the equilibrium distribution $ p_{\alpha}[equi] $ exists, it satisfies the equation,
\begin{equation} 
p_{\alpha} [equi] w_{ \alpha \rightarrow \beta} = p_{\beta} [equi] w_{\beta \rightarrow \alpha}, 
\end{equation}
known as the detailed balance.
From one of the axioms, which states that all the possible worlds are equally probable \footnote{We shall call it the equal probability axiom.}, $ p_{\alpha}[equi] $ should be equal for any distinct configuration $ \alpha $.

The equal probability axiom results the state independence of the transition probability $ w_{ \alpha \rightarrow \beta} $ between neighboring two configurations, and the master equation can be rewritten within the reduced space of the first two quantum numbers, $ a = ( N_2, \tilde{N}_1 ) $, as 
\begin{equation} 
\tilde{p}_a [n+1]= \tilde{p}_a [n]+ \sum_b \{\tilde{p}_b [n] \tilde{w}_{b \rightarrow a}-\tilde{p}_a [n] \tilde{w}_{a \rightarrow b} \}, 
\end{equation}
where the sum over $ b $ runs from 1 to 4 corresponding to the 4 moves, which shift $\{N_2,\tilde{N}_1\}$ to $ \{ N_2 \pm 1, \tilde {N}_1 \pm 1 \} $.
The detailed balance condition in the reduced space is then written as,
\begin{equation}
\tilde{p}_a [equi] \tilde{w}_{a \rightarrow b} = \tilde{p}_b [equi] \tilde{w}_{b \rightarrow a}. 
\end{equation}
As a consequence of the equal probability axiom, $ \tilde{p}_a [equi] $ should be proportional to the number of configurations, $ \phi_a $.
The detailed balance condition can be satisfied by taking the transition probability $ \tilde{w}_{a \rightarrow b} $ as
\begin{equation}
\tilde{w}_{a \rightarrow b} = \frac{\phi_b}{\phi_a + \phi_b}. \label{eq.2.5}
\end{equation}
It may be possible to choose another form for $ \tilde{w}_{a \rightarrow b} $ which fulfill the detailed balance condition. 
Our specific choice of $\tilde{w}_{a \rightarrow b}$ will be justified when we take the continuous limit of the master equation.

The master equation in the continuum space is obtained by taking the limit that the scale parameters $ \{ A_0, l_0, \tau_0 \} $, introduced for the area $ A = A_0 N_2 $, the boundary length, $ l = l_0 \tilde {N}_1 $, and the Markov process time\footnote{We may call it simply time unless we need to distinguish it from the physical time which will be defined later.}, 
$ \tau = \tau_0 n$, tend to zero.
Leaving details of the derivation in Appendix A, it is written as
\begin{equation}
\frac{\partial}{\partial \tau} p[A,l;\tau] =\Bigl\{ \Bigl(\frac{\partial^2}{\partial A^2} + \frac{\partial}{\partial A} \frac{\partial S_a}{\partial A} \Bigr) + \Bigl( \frac{\partial^2}{\partial l^2}+ \frac{\partial}{\partial l} \frac{\partial S_a}{\partial l} \Bigr ) \Bigr\} p[A,l;\tau], 
\label{eq.2.6}
\end{equation}
where $ S_a \equiv S(A,l) $ is defined by $ \phi_a = \exp(-S_a) $.
Scale parameters are tuned as 
$$
\frac{A_0^2}{ \tau_0}  = \frac{l_0^2}{ \tau_0} .
$$
Eq.(2.6) clearly shows that the distribution $ \phi_a $ is the stationary solution, and it is the reason of our choice for the transition probability, $ \tilde{w}_{a \rightarrow b} $.
The function $ \phi_a $ was computed numerically by the dynamical triangulation with a boundary\cite{Adi}, and later the exponent was analytically given by the boundary Loiuville theory\cite{FZZ} and the matrix theory\cite{HY} as
\begin{equation}
S(A,l) = - \alpha \log A - \beta \log l + \gamma \frac{l^2}{A}  + \mu A + \mu^B l, \label{eq.2.7}
\end{equation} 
where $ \alpha = -1-1/b^2 $, $\beta = -2+1/b^2$, and $ \gamma = 1/(4 \sin \pi b^2) $.
The parameter $ b $ is related to the Liouville background charge $ Q $ through $ Q = b+1/b $.
In order to avoid the singular behavior of the master equation at those $ b $ where $\gamma$ diverges, we rescale the area, the boundary length, and the time by $ \gamma $, as $ A = x/\gamma$, $ l = y/\gamma $, and $ \tau=t/\gamma^2 $, respectively, which leads
\begin{equation}
\frac{\partial}{\partial t} p[x,y;t] = \Bigl\{\frac{\partial^2}{\partial x^2} + \frac{\partial}{\partial x} \frac{\partial S(x,y)}{\partial x} + \frac{\partial^2}{\partial y^2} + \frac{\partial}{\partial y} \frac{\partial S(x,y)}{\partial y}\Bigr\} p[x,y;t], \label{eq.2.8}
\end{equation} 
with
\begin{equation}
S(x,y) = - \alpha \log x - \beta \log y + \frac{y^2}{x} 
 + \lambda x + \lambda^B y + 3 \log \gamma. \label{eq.2.9}
\end{equation} 
The cosmological constant, $\lambda$, and the boundary cosmological constant, $\lambda^B$, are also rescaled accordingly as $ \mu = \gamma \lambda $ and $ \mu^B = \gamma \lambda^B $.
The last term in $ S(x, y) $ modifies the overall normalization of the distribution, and we shall put this term in the normalization constant.

\begin{figure}[ht]
\begin{center}
\includegraphics{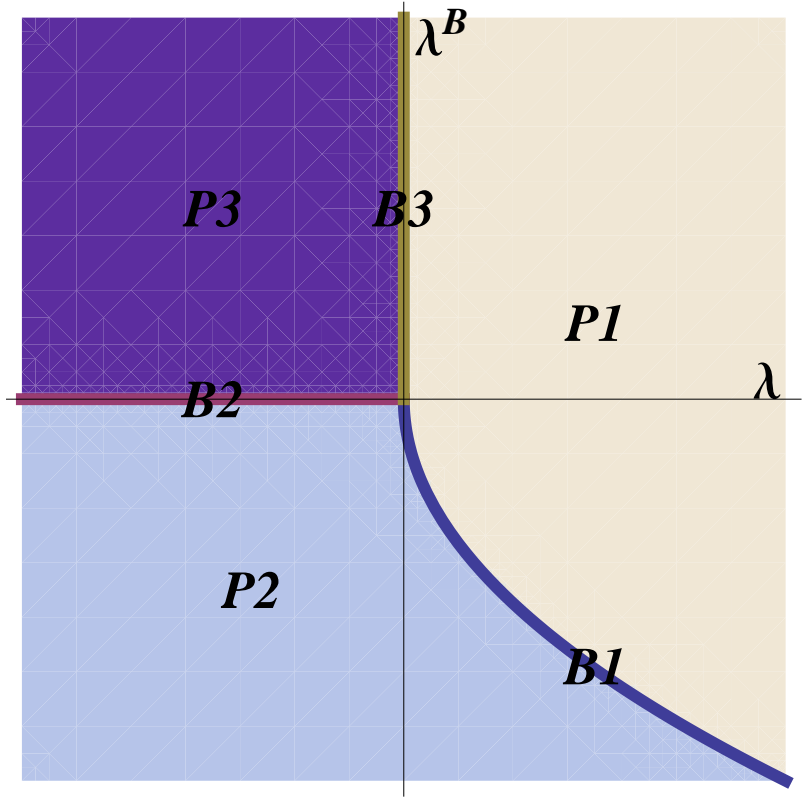}
\end{center}
\caption{\small Collapsing phase({\it P1}). Expanding phase({\it P2}). Elongating phase({\it P3}). The boundary state is on the boundary {\it B2}, and the Polymer state is in the elongating phase {\it P3}.\label{phases}}
\end{figure}

In the two dimensional parameter space ($ \lambda$, $\lambda^B $) the numerical simulation for the pure gravity ($ 1/b^2 = 3/2 $) has revealed three critical lines separating into regions corresponding to three distinguished phases of universe.
We shall write each phase as {\it P1}, {\it P2}, and  {\it P3}, and name them as the collapsing, the expanding, and the elongating phase, respectively.
We also write the boundaries between {\it P1} and {\it P2} as {\it B1}, {\it P2} and  {\it P3} as {\it B2}, and  {\it P1} and  {\it P3} as {\it B3}.
The relative locations of phases and boundaries are depicted in Figure\ref{phases}.
%
%
The numerical simulation has shown that the area tends to decrease towards a point universe when the parameters ($ \lambda, \lambda^B $) are in the region of the collapsing phase, which locates most of the right half-plane of the parameter space, while the area grows indefinitely with the maximum rate when the parameters are outside of {\it P1}. 
This non-static region is divided into two phases by the line $\lambda^B=0$ in the $\lambda<0$ half-plain.
Above the critical line the boundary length tends to a constant either like an elongating tube or string with the minimum radius. 
Below the critical line both the area and the boundary length increase with the maximum rates.

In the subsequent sections we shall obtain possible solutions for various phases, which are classified according to the behavior of average area and boundary length in the parameter space ($\lambda$, $\lambda^B$, $\beta$).

\section{The collapsing phase ({\it P1}~)}
\setcounter{equation}{0}
\noindent
We begin with studying the static solution $ \tilde{p}_a [equi] \sim \exp (-S_a) $ of the master equation.
Integrals in the generating function,
\begin{equation}
Z[ \lambda, \lambda^B] = \int \exp\{-S(x,y)\} dx dy, \label{eq.3.1}
\end{equation} 
are in general diverging, and we need to regularize it.
We regard it as the continuum limit of the matrix model partition function,
\begin{equation}
Z_M[ \mu, \mu^B]=\sum_{( N_2,\tilde {N}_1 )} \exp(-\mu N_2 - \mu^B \tilde {N}_1)~ \tilde{Z}_M [N_2, \tilde {N}_1], \label{eq.3.2}
\end{equation}
where the sums over $ ( N_2,\tilde{N}_1 ) $ are restricted by $ N_2 \geq \tilde{N}_1-2, \tilde {N}_1 \geq 3 $, and $ N_2+\tilde{N}_1=$ {\it even integer}, due to the manifold conditions we have employed.

The number of distinct congigurations has been calculated either by the graph theory\cite{TUTTE} or the matrix model\cite{MATRIX-THEORY}, which approaches in the large $N_2$ limit\cite{HY} as 
\begin{equation}
\tilde {Z}_M [N_2,\tilde {N}_1] \sim N_2^{\alpha_M} \tilde {N}_1^{\beta_M} 
\exp \Bigl( -\gamma_M  \frac{\tilde{N}_1^2}{N_2} \Bigr) \exp( \mu_c N_2 + \mu^B_c  \tilde{N}_1 ) \label{eq.3.3},
\end{equation}
where the exponents are given for the pure gravity by $ \alpha_M = -5/2 $ and $ \beta_M = 1/2 $. 
The non-universal constants $ \gamma_M $, $ \mu_c $, and $ \mu^B_c $ depend on the choice of manifold conditions employed for the model according to the types of graphs or diagrams included.
The matrix model partition function is then written as
\begin{eqnarray}
Z_M[\mu, \mu^B] 
&\sim& \sum_{N_2 \geq 1} \sum_{\tilde {N}_1 \geq 3} 
N_2^{\alpha_M} \tilde {N}_1^{\beta_M} 
\exp \Bigl(-\gamma_M  \frac{\tilde{N}_1^2}{N_2}\Bigr) \nonumber \\  
&\times& \exp \{-(\mu-\mu_c) N_2 -(\mu^B-\mu^B_c) \tilde{N}_1\}. \label{eq.3.4}
\end{eqnarray}
Here, we note the constraints in the sums over $ (N_2, \tilde{N}_1) $ are slightly relaxed owing to the strong damping factor $ \exp ( -\gamma_M \tilde{N}_1^2/N_2 ) $.
Sums in the partition function are approximated to integrals in the continuum limit by making scale factors for the area and the boundary length, $ A_0 $, and $ l_0 $, tend to zero:
\begin{equation}
Z[\lambda,\lambda^B] \rightarrow  \int_{x_0}^{x_1} dx \int_{y_0}^{y_1} dy~  x^{\alpha} y^{\beta} \exp\Bigl( - \frac{y^2}{x} \Bigr) \exp(-\lambda x-\lambda^B y), \label{eq.3.5}
\end{equation}
where we keep the regions of integrals finite in order to study behaviors toward the continuum limit, $ (x_0, y_0) \rightarrow 0 $, as well as the large size limit, $ (x_1, y_1) \rightarrow \infty $.
The cosmological constants are shifted and rescaled accordingly as $ \lambda \equiv (\mu-\mu_c) / (\gamma A_0) $, and $ \lambda^B \equiv (\mu^B-\mu^B_c) / (\gamma l_0) $. 

We generalize the exponents $ (\alpha, \beta) $ through employing values obtained by the boundary Liouville theory, $ \alpha = -1-1/b^2 $ and $ \beta = -2 + 1/b^2 $, in order to take into account matter fields, which are assumed to couple conformally with the space-time. 
We should remark that the difference in exponents $ \beta_M =1/2$ and $ \beta=-1/2 $ for the pure gravity is due to the way of counting distinct configurations. 
The matrix model respects the point group multiplicity factor, $ \tilde{N}_1$, in the state number counting.
On the other hand the boundary Liouville theory regards those configurations which are members of the same point group symmetry are identical, since the universe stands by itself and does not have any reference frame.
We shall employ the latter viewpoint in the followings.

The integration over $ x $ converges, both in the limit $ x_0 \rightarrow 0 $ due to the factor $ \exp (-y^2/x) $ for $ y \geq y_0 >0 $, and in the limit $ x_1 \rightarrow \infty $ because of $ \exp (-\lambda x) $ for $ \lambda>0 $, to
\begin{equation}
\int_0^{\infty} dx~ x^{\alpha} \exp \Bigl(- \frac{y^2}{x}\Bigr) \exp (-\lambda x) 
= 2 \lambda^{\nu/2} y^{-\nu} K_{\nu}[2 \sqrt{\lambda} y], \label{eq.3.6}
\end{equation}
where we define $ \nu \equiv 1/b^2 = -(1+\alpha) $, and where $ K_{\nu}[z]$ is the modified Bessel function.
We note that according to the small $z$ behavior of the Bessel function, $ K_{\nu}[z] \sim z^{-\nu}$, the integral converges even at $\lambda=0$, if we recall $\nu \geq 1$.
On the other hand the integration over $ y $,
$$
\int_{y_0}^{y_1} dy~ y^{\beta} \exp (-\lambda^B y) y^{-\nu} 
K_{\nu}[2 \sqrt{\lambda} y],
$$
requires a UV cut-off, $y_0$, while in large $ y $ regions it is possible to take the limit $ y_1 \rightarrow \infty $, if $ 2 \sqrt{\lambda} + \lambda^B > 0 $, since asymptotic form of the integrand contains the exponential factor $\exp\{-(2 \sqrt{\lambda} + \lambda^B) y \}$.  
The critical line separating the static phase and non-static phases locates along $  \lambda^B+2 \sqrt{\lambda} = 0 $ in the lower half-plane of the parameter space $ (\lambda, \lambda^B) $, which we named as {\it B1}, and along the line $ \lambda=0 $ in the upper half plane  as {\it B3}.

Since all the statistical moments we are interested in are obtained by making use of the relation 
\begin{equation}
 \langle x^n y^m\rangle =\frac{(-1)^{n+m}}{Z[\lambda,\lambda^B]} \frac{\partial^n}{\partial \lambda^n} \frac{\partial^m}{(\partial \lambda^B)^m} Z[\lambda,\lambda^B] ,\label{3.7}
\end{equation} 
where the statistical average is represented by brackets $\langle\dots\rangle$, we shall pay our attention on diverging parts of the integral which dominate in the continuum limit, $ y_0 \rightarrow 0 $.
Utilizing the power series expansion of the modified Bessel function $K_{\nu}[2 \sqrt{\lambda} y]$, and expanding the exponential factor $ \exp (- \lambda^B y) $, we obtain the diverging parts of the integral as 
\begin{equation}
Z[\lambda,\lambda^B] = \sum_{n,m} \frac{(-1)^{n+m}}{n!m!} \lambda^n (\lambda^B)^m Z_{(n,m)}, \label{3.8}
\end{equation}
with
\begin{equation}
Z_{(n,m)}\sim 
\left\{ \begin{array}{ll}
     \Gamma ( \nu-n )\frac{y_0^{2n+m-1-\nu}}{\nu+1-(2n+m)} &\mbox{if}\ \nu>2n+m-1 \\
     0 &\mbox{if}\ \nu<2n+m-1,
      
        \end{array}  
\right.  \label{eq.3.9} \\
\end{equation}
for $\nu>1$, and for $\nu=1$ we have
\begin{eqnarray}
Z[\lambda,\lambda^B]&\sim& \frac{1}{2 y_0^2} - \frac{\lambda^B}{y_0} \nonumber \\
&-&\frac{1}{2} \{(\lambda^B)^2-2 \lambda+4\gamma_E \lambda+2 \lambda \log \lambda+2 \lambda \log y_0\}\log y_0,\label{eq.3.10}
\end{eqnarray}
where $\gamma_E$ represents the Euler gamma.

From the partition function we can calculate statistical averages $ \langle x\rangle $, $ \langle y\rangle $, and correlation functions $ \langle\Delta_x^n \Delta_y^m\rangle $, of fluctuations, $ \Delta_x \equiv x-\langle x \rangle $ and $ \Delta_y \equiv y-\langle y \rangle $, for any pair of non-negative integers $ (n,m) $. 
For example, $ \langle x\rangle $ and $ \langle y\rangle $ are given by
\begin{eqnarray}
\langle x\rangle &=& -\frac{\partial}{\partial \lambda} \log Z \sim -\frac{Z_{(10)}}{ Z_{(00)}} \nonumber \\
                 &=& \left\{
                       \begin{array}{ll}
                         \frac{ \nu+1 }{ (\nu-1)^2 } y_0^2, & \mbox{if}\ \nu > 1 \\
                         2 (y_0 \log y_0)^2  & \mbox{if}\ \nu=1,\\
                       \end{array}
                     \right.   \\
\langle y\rangle &=& -\frac{\partial}{\partial \lambda^B} \log Z \sim -\frac{Z_{(01)}}{ Z_{(00)}} \nonumber \\
                 &=& \frac{ \nu+1 }{ \nu } y_0 . 
                           \label{eq.3.12}
\end{eqnarray}
They are independent of the parameters $\lambda$ and $\lambda^B$, and the collapsing phase approaches smoothly toward the {\it B1} and {\it B3} boundaries.
Since the average values $ \langle A\rangle = \langle x\rangle/\gamma$ and $ \langle l\rangle =\langle y\rangle /\gamma$ vanish in the continuum limit $ y_0 \rightarrow 0 $, it is the reason we call this phase as collapsing.

The second moment, $ \langle\Delta_x^2\rangle $ is given by
\begin{eqnarray}
\langle\Delta_x^2\rangle &=& \frac{\partial^2}{\partial \lambda^2} \log Z \sim \frac{Z_{(20)}}{Z_{(00)}} - \left\{\frac{Z_{(10)}}{Z_{(00)}}\right\}^2 \nonumber \\
                         &=& \left\{
                           \begin{array}{ll}
                              \{\frac{1}{(\nu-2)(\nu-3)}-      
                      \frac{\nu+1}{(\nu-1)^3}\}\frac{\nu+1}{\nu-1} y_0^4 & \mbox{if}\ \nu>3 \\
                              -2 y_0^4 \log y_0 & \mbox{if}\ \nu=3 \\
                              0 & \mbox{if}\ 1 \leq \nu <3,  \label{eq.3.13}
                           \end{array}
                              \right.
\end{eqnarray}
where the dominant contribution of the last case will come from the non-diverging integral.
In the same manner we can obtain other second order moments as,
\begin{eqnarray}
\langle\Delta_x \Delta_y\rangle &=& \frac{\partial^2}{\partial \lambda \partial \lambda^B } \log Z \sim \frac{Z_{(11)}}{Z_{(00)}}-\frac{Z_{(10)} Z_{(01)}}{Z_{(00)} ^2} \nonumber \\
                                &=& \left\{
                                  \begin{array}{ll}
                                     \{\frac{1}{\nu-2}-   
                                     \frac{\nu+1}{\nu(\nu-1)}\}\frac{\nu+1}{\nu-1} y_0^3 & 
                                     \mbox{if}\ \nu>2 \\
                                     -3 y_0^3 \log y_0 & \mbox{if}\ \nu=2 \\
                                     0 & \mbox{if}\ 1 \leq \nu <2, 
                                  \end{array}
\right. \\
       \langle\Delta_y^2\rangle &=& \frac{\partial^2}{(\partial\lambda^B)^2} \log Z \sim \frac{Z_{(02)}}{Z_{(00)}}- \Bigl \{\frac{Z_{(01)}}{Z_{(00)}}\Bigr \}^2 \nonumber \\
                                &=& \left\{
                                      \begin{array}{ll} 
                                        (\frac{1}{\nu-1}- \frac{\nu+1}{ \nu^2})(\nu+1) y_0^2 & \mbox{if}\ \nu>1 \\
 - y_0^2 \log y_0 & \mbox{if}\ \nu=1. \label{eq.3.15}
      \end{array}
    \right.
\end{eqnarray}
In the continuum limit all the correlation functions also vanish as
\begin{equation}
\langle \Delta_x^n \Delta_y^m \rangle \sim 
\left\{
  \begin{array}{ll}
{\cal O} (y_0^{2n+m}) & \mbox{if}\ \nu>2n+m-1\\
 -y_0^{2n+m} \log y_0 & \mbox{if}\ \nu=2n+m-1\\
 {\cal O} (y_0^{\nu+1}) & \mbox{if}\ 1 \leq \nu <2n+m-1. \label{eq.3.16}
  \end{array}
\right.
\end{equation}

\section{Overview of non-stationary solutions of the master equation}

\setcounter{equation}{0}
\noindent
In order to investigate solutions of the master equation for non-stationary phases we shall analyze time evolution of statistical averages of moments $\langle x^n y^m \rangle$, instead of solving the partial differential equation.
The master equation provides equations of the motion as
\begin{eqnarray}
\frac{d \langle x^n y^m \rangle} {dt} &=& n(n-1) \langle x^{n-2} y^m\rangle+ m(m-1) \langle x^n y^{m-2}\rangle \nonumber \\
&-& n \langle x^{n-1} y^m \frac{\partial S}{\partial x}\rangle - m \langle x^n y^{m-1} \frac{\partial S}{\partial y}\rangle, \label{eq.4.1}
\end{eqnarray}
where $ S=S(x,y) $ has been given by eq.(2.9).
When the parameters ($\lambda$, $\lambda^B$, $\beta$) are chosen outside the region where the solution converges to the equilibrium distribution, it is beyond the case where standard procedures of the statistical mechanics are applicable.
We shall require the master equation with the same function $S(x,y)$ holds in the whole parameter space because of the equal probability axiom.

Equations of the motion for $\langle x\rangle$ and $\langle y\rangle$ are given by
\begin{eqnarray}
\frac{d \langle x \rangle} {d t} &=& - \lambda + \langle\frac{y^2}{x^2}\rangle + \alpha \langle\frac{1}{x}\rangle, \\
\frac{d \langle y \rangle} {d t} &=& - \lambda^B -2 \langle\frac{y}{x}\rangle + \beta \langle\frac{1}{y}\rangle.\label{eq.4.3}
\end{eqnarray}
Since $x$ and $y$ correspond to the rescaled area and the boundary length, average values $\langle y^2/x^2\rangle$ and $\langle y/x\rangle$ should be positive definite, even if $\langle x\rangle$ and $\langle y\rangle$ grow to infinity.
Assuming $\langle y^2/x^2\rangle \rightarrow c_2$ and $\langle y/x\rangle \rightarrow c_1$, eq.(4.3) can be written asymptotically as\footnote{In the following 'tilde'($\sim$) means asymptotically equal.}
\begin{eqnarray}
\frac{d \langle x \rangle} {d t} &\sim& - \lambda + c_2 + \alpha \langle\frac{1}{x}\rangle, \\ 
\frac{d \langle y \rangle} {d t} &\sim& - \lambda^B - 2 c_1 + \beta \langle\frac{1}{y}\rangle.
\end{eqnarray}
In these equations $\langle1/x\rangle$ and $\langle1/y\rangle$ are expected to behave like $1/\langle x\rangle$ and $1/\langle y\rangle$, respectively, when $\langle x\rangle$ and $\langle y\rangle$ get large.
On the other hand, they will increase monotonically as $\langle x\rangle$ and $\langle y\rangle$ move toward zeros.
Recalling the parameter $ \alpha $ is negative, $\langle x\rangle$ increases indefinitely when $-\lambda + c_2$ is positive, while it goes to zero otherwise.\footnote{Precisely speaking, because of the repulsive nature of the fixed point located at $-\lambda+c_2+\alpha\langle1/x\rangle=0$, the distribution bifurcates into the collapsing phase and the expanding phase with appropriate ratio, which tends 100\% collapsing as $-\lambda+ c_2$ approaches to $0$.}

Therefore, when $\langle x\rangle \rightarrow 0$, $\langle y\rangle$ should also tend to zero, which is the case we have already seen as the collapsing phase.
On the other hand, if $\langle x\rangle \rightarrow \infty$, the increasing rate should be at most proportional to the Markov time.
In fact, eq.(4.2) gives the asymptotic solution $\langle x\rangle\sim (-\lambda+c_2) t$.

The evolution of $\langle y\rangle$ depends on the parameters $\lambda^B$ and $\beta$.
When $\beta$ is negative, $ d \langle y \rangle/d t $ is negative and the solution becomes unstable except $\beta=-1$, which will be discussed in the later section as the polymer state.
On the other hand when $\beta$ is positive, $\langle y\rangle$ converges to an attractive fixed point for $- \lambda^B-2 c_1$ being negative, otherwise it expands with the maximum rate, as far as the parameter $\lambda$ lies outside the region of the collapsing phase.

In summary, when $\lambda^B < -2 c_1$ and $\lambda < c_2$ , both $\langle x\rangle$ and $\langle y\rangle$ increase linearly in time, which corresponds to the expanding phase({\it P2}~).
Asymptotically $c_2$ and $c_1$ tend to $\langle y\rangle^2/\langle x\rangle^2$ and $\langle y\rangle/\langle x\rangle$, respectively, and the critical line is located along $\sqrt{\lambda}=-2 \lambda^B$ in the parameter space.
The rest of the lower half-plane of the parameter space is occupied by the collapsing phase({\it P1}~).
In the upper half-plane of the parameter space $\langle y\rangle$ converges to a fixed point, and $c_2$ and $c_1$ go to zeros when $\lambda\rightarrow 0$.
This is the elongating phase({\it P3}~), while the remaining region in the upper half-plane corresponds to the collapsing phase.

These three phases are separated by three boundaries:the boundary 
{\it B1} between {\it P1} and {\it P2} is located along the line $\sqrt{\lambda}= -2\lambda^B$ in the fourth quadrant.
Since $d\langle x\rangle /dt$ is negative on this border, no physical solution exist, and the transition between two phases is abrupt.
the boundary {\it B2}, which lies between {\it P2} and {\it P3}, is specified by $-\lambda > 0$ and $\lambda^B=0$.
Since $d\langle x\rangle /dt=-\lambda$ and $d\langle y\rangle /dt=\beta\langle1/y \rangle$, there exists a solution where the space expands as $\langle x \rangle\sim {\cal O}(t)$ and $\langle y \rangle\sim {\cal O}(t^{1/2})$, which we call the boundary state.
The boundary {\it B3} lies between {\it P3} and {\it P1}, where $\lambda=0$ and $\lambda^B > 0$.
Since $d\langle x\rangle /dt$ is also negative on this border, no physical solution exist, and the transition between two phases is sudden similar to {\it B1}.

In the following sections we shall show explicit forms of asymptotic solutions for these phases and boundaries.

\section{The expanding phase ({\it P2}~)}
\setcounter{equation}{0}
\noindent
When we choose the parameters $ \lambda $ and $ \lambda^B $ negative, configurations with more triangles and more boundary edges become favorable, and the universe expands with its maximum growth rates proportional to the Markov time. 
We have called this case as the expanding phase after the numerical simulation.
In such a case it is common to separate fluctuations from their means as $ x = \langle x \rangle + \Delta_x $ and $y = \langle y \rangle + \Delta_y $.
Equations of the motion are constructed for the averages $\langle x \rangle$, $\langle y \rangle$, and various moments of fluctuations $\langle\Delta_x^n \Delta_y^m \rangle$, which are summarized in the Appendix B.
Since the mean values, $ \langle x \rangle $ and $ \langle y \rangle $, increase with their maximum rates, we can expand equations of the motion in terms of the relative fluctuations, $ \xi \equiv \Delta_x/\langle x \rangle $ and $ \eta \equiv \Delta_y/\langle y \rangle $. 
As we shall see in a moment, the mean square fluctuations, $ \langle\Delta_x^2 \rangle $ and $ \langle\Delta_y^2 \rangle $, also show linear increase in time, the order parameters of expansions for $ x $ and $ y $ motions, $ \sqrt{\langle\xi^2 \rangle} $ and $ \sqrt{\langle\eta^2 \rangle} $, are proportional to $ t^{-1/2} $.
Therefore, solutions obtained by keeping the leading orders of expansion become exact asymptotically.

For example, asymptotic equations for $ \langle x \rangle $ and $ \langle y \rangle $ are obtained from eq.(4.2) and (4.3) as 
\begin{eqnarray}
\frac{d \langle x \rangle}{d t} &\sim& - \lambda + \frac{\langle y \rangle^2}{\langle x \rangle^2} \\
\frac{d \langle y \rangle}{d t} &\sim& - \lambda^B - 2 \frac{\langle y \rangle}{\langle x \rangle}. \label{eq.5.1}
\end{eqnarray}
Solutions of these equations are easily obtained by writing 
$$
 \langle x \rangle \sim v_x t , ~~\langle y \rangle \sim v_y t , 
$$
where we keep only the leading terms, and require that $ v_x $ and $ v_y $ should satisfy
\begin{eqnarray}
v_x &=& -\lambda + \frac{v_y^2}{v_x^2},\\
v_y &=& -\lambda^B -2 \frac{v_y}{v_x}. \label{eq.5.2}
\end{eqnarray}
The positivity of growth rates, $ v_x $ and $ v_y $, defines two boundaries of the expansion phase in the 2-dimensional parameter space.
The conditions, $ v_x=0 $ and $ v_y =0 $, with keeping the ratio $ v_y/v_x(\equiv r_0) $ finite, give the boundary {\it B1} along $ 4 \lambda = (\lambda^B)^2 $ with $ \lambda^B <0 $.
This boundary coincides with that of the collapsing phase in the lower half-plane as we have seen in Sec.3.
There exists another boundary {\it B2}, where $ v_x= - \lambda $ and $ v_y =0 $, which lies along $ \lambda^B = 0 $ with $ \lambda <0 $. 

Asymptotic equations for the 2-point correlation functions are written as
\begin{eqnarray}
\frac{d \langle\Delta_x^2 \rangle}{d t} &\sim& 2 - 4  \frac{\langle y \rangle^2}{\langle x \rangle^2} \Bigl( \frac{\langle\Delta_x^2 \rangle}{\langle x \rangle} -  \frac{\langle \Delta_x \Delta_y \rangle}{\langle y \rangle}\Bigr), \\
\frac{d\langle\Delta_x \Delta_y \rangle}{dt} &\sim& - 2 \frac{\langle y \rangle^2}{\langle x \rangle^2}\Bigl( \frac{\langle\Delta_y^2 \rangle}{\langle y \rangle} - \frac{\langle \Delta_x \Delta_y \rangle}{\langle x \rangle}\Bigr) \nonumber \\
& &~ + 2 \frac{\langle y \rangle}{\langle x \rangle}\Bigl( \frac{\langle\Delta_x^2 \rangle}{\langle x \rangle} - \frac{\langle \Delta_x \Delta_y \rangle}{\langle y \rangle}\Bigr), \\
\frac{d \langle\Delta_y^2 \rangle}{d t}  &\sim& 2 - 4 \frac{\langle y \rangle}{\langle x \rangle} \Bigl( \frac{\langle\Delta_y^2 \rangle}{\langle y \rangle} - \frac{\langle \Delta_x \Delta_y \rangle}{\langle x \rangle}\Bigr) . \label{eq.5.3}
\end{eqnarray}
They also grow linearly in $ t $. Writing 
$$ 
\langle \Delta_x^2 \rangle \sim c(2,0)t, ~~ \langle \Delta_x \Delta_y \rangle \sim c(1,1)t ,  ~~\langle \Delta_y^2 \rangle \sim c(0,2)t. 
$$
their coefficients are obtained to be 
\begin{eqnarray}
c(2, 0) &=& \frac{2 (4 + v_x)}{4 + 4 r_0^2 + v_x},\nonumber \\
c(1, 1) &=& \frac{8 r_0}{4 + 4 r_0^2 + v_x}, \\
c(0, 2) &=& \frac{2 (4 r_0 + v_x)}{4 + 4 r_0^2 + v_x}.\nonumber \label{eq.5.4}
\end{eqnarray}

The asymptotic equation for the $(n,m)$-moment is separated into two branches, $ n + m = $ {\it even} and $ n + m = $ {\it odd}.
For $ n + m = N$ with an even integer $N$ there are $N+1$ equations,
\begin{eqnarray}
\frac{d\langle\Delta_x^n \Delta_y^m \rangle}{d t} &\sim& n(n-1) \langle\Delta_x^{n-2} \Delta_y^ m \rangle+ m(m-1) \langle\Delta_x^n \Delta_y^{m-2}\rangle \nonumber \\
& &~-2n \frac{\langle y \rangle^2}{\langle x \rangle^2} \Bigl(\frac{\langle\Delta_x^{n} \Delta_y^m \rangle}{\langle x \rangle} - \frac{\langle\Delta_x^{n-1} \Delta_y^{m+1}\rangle}{\langle y \rangle}\Bigr) \nonumber \\
& &~+2m \frac{\langle y \rangle}{\langle x \rangle} \Bigl(\frac{\langle\Delta_x^{n+1} \Delta_y^{m-1}\rangle}{\langle x \rangle} - \frac{\langle\Delta_x^n \Delta_y^m \rangle}{\langle y \rangle}\Bigr).\label{eq.5.5}
\end{eqnarray}
Inserting the asymptotic form of the solution, 
$$
\langle\Delta_x^ {n} \Delta_y^ {m}\rangle \sim c(n,m)  t^{\frac{n+m}{2}},
$$
into eq.(5.9), we obtain
\begin{eqnarray}
\Bigl(\frac{n+m}{2}&+&\frac{2 n r_0^2}{v_x} + \frac{2 m}{v_x}\Bigr) c(n,m)\nonumber \\
&=& n(n-1)c(n-2,m)+ m(m-1) c(n,m-2) \nonumber \\
& &+ \frac{2n r_0}{v_x} c(n-1,m+1) + \frac{2mr_0}{v_x} c(n+1,m-1), \label{eq.5.6}
\end{eqnarray}
which can be solved step by step starting with the $ n + m = 2 $ components with the initial input $ c(0,0) = 1 $.

For the $ n + m =${\it odd integer} branch the equation has extra contributions from $ n + m =${\it even integer} components,
\begin{eqnarray}
 & & n\frac{\langle y \rangle^2}{\langle x \rangle^2} \Bigl( 3\frac{\langle\Delta_x^{n+1} \Delta_y^m \rangle - \langle\Delta_x^ {n-1} \Delta_y^m \rangle\langle\Delta_x^2 \rangle}{\langle x \rangle^2} \nonumber \\
&-& 4\frac{\langle\Delta_x^n \Delta_y^{m+1}\rangle - \langle\Delta_x^ {n-1}\Delta_y^m \rangle\langle\Delta_x\Delta_y \rangle}{\langle x \rangle\langle y \rangle} \nonumber \\
&+& \frac{\langle\Delta_x^{n-1}\Delta_y^ {m+2}\rangle - \langle\Delta_x^{n-1}\Delta_y^m \rangle\langle\Delta_y^2 \rangle}{\langle y \rangle^2}\Bigr) \nonumber \\ 
&-& 2m\frac{\langle y \rangle}{\langle x \rangle} \Bigl(\frac{\langle\Delta_x^{n+2}\Delta_y^{m-1}\rangle - \langle\Delta_x^n \Delta_y^{m-1}\rangle\langle\Delta_x^2 \rangle}{\langle x \rangle^2} \nonumber \\
&-& \frac{\langle\Delta_x^{n+1}\Delta_y^m \rangle - \langle\Delta_x^n \Delta_y^{m-1}\rangle\langle\Delta_x \Delta_y \rangle}{\langle x \rangle\langle y \rangle}\Bigr), 
\end{eqnarray}
which should be added to eq.(5.9).
Inserting the expected asymptotic form,
$$
 \langle\Delta_x^ {n} \Delta_y^ {m}\rangle  \sim  c(n,m)  t^{\frac{n+m-1}{2}} ,
$$ 
into eq.(5.9) with eq.(5.11) we obtain 
\begin{eqnarray}
\Bigl(\frac{n+m-1}{2}&+&\frac{2 n r_0^2}{v_x} + \frac{2 m}{v_x}\Bigr) c(n,m) \nonumber \\
&=& n(n-1)c(n-2,m)+ m(m-1) c(n,m-2) \nonumber \\
& &+ \frac{2n r_0}{v_x} c(n-1,m+1) + \frac{2m r_0}{v_x} c(n+1,m-1)\nonumber \\
& &+ \frac{n}{v_x^2} [3 r_0^2 \{ c(n+1,m)-c(n-1,m) c(2,0) \} \nonumber \\
& &- 4 r_0 \{ c(n,m+1)-c(n-1,m) c(1,1) \} \nonumber \\
& &+ c(n-1,m+2)-c(n-1,m) c(0,2)] \nonumber \\
& &-\frac{2m}{v_x^2} [r_0 \{ c(n+2,m-1)-c(n,m-1)c(2,0) \} \nonumber \\
& &- c(n+1,m)+c(n,m-1)c(1,1)], \label{eq.5.7}
\end{eqnarray}
This equation can also be solved iteratively starting from $ n + m = 3 $, with initial inputs $ c(1,0) = c(0,1) = 0 $, together with $ n + m = even $ solutions previously obtained.

Near the boundary {\it B1} of the expanding phase there are two cases of limiting procedures: {\it 1}) take the $t \rightarrow \infty$ limit before taking $(v_x, v_y) \rightarrow 0$, or {\it 2}) search a new state just on the boundary before taking the asymptotic limit.
In the first case the expansion in terms of the relative fluctuations, $\xi$ and $\eta$, is considered to be valid.
As the parameters get close to the boundary values, the leading terms of expansion rates for $ \langle x \rangle $ and $ \langle y \rangle $ tend to vanish as 
$$
v_x \sim \frac{\sqrt{\lambda}}{1+\lambda} \delta,
~~v_y \sim \frac{\lambda}{1+\lambda} \delta ,
$$
for a small $ \delta \equiv -(\lambda^B + 2 \sqrt{\lambda}) $.
On the other hand the growing rates for second moments stay finite towards {\it B1} seen in eq.(5.8) for $ v_x \rightarrow 0$ .
Higher moments of correlation are obtained by solving eq.(5.10) and eq.(5.12), and then take the $v_x \rightarrow 0$ limit.

For the second limiting case we search the following type of solutions on the boundary:
$$
\langle x \rangle \sim x_1 t^{p} + \dots,~~\langle y \rangle \sim y_1 t^{p} + \dots, 
$$
where $0<p<1$.
The correlation functions we look for are considered to have the form, 
\begin{equation}
\langle\Delta_x^n \Delta_y^m \rangle \sim 
\left\{
\begin{array}{ll}
 c_1(n,m)t^{\frac{n+m}{2}+q-1} &\mbox{if} \ n+m={\it even}  \\
 c_1(n,m)t^{\frac{n+m-1}{2}+q-1} &\mbox{if} \ n+m={\it odd}
\end{array}
\right.
\end{equation}
with $0<q \leq 1$.
At this stage we assume equations of the motion can be expanded in terms of $ \xi $ and $ \eta $, and we take up to the second order, whose validity will be checked later.
Inserting the asymptotic forms, equations for the averages $ \langle x \rangle $ and $ \langle y \rangle $ are given by
\begin{eqnarray}
p x_1 t^{p-1} &\sim& \frac{ \alpha }{x_1} t^{-p} \Bigl(1 +\frac{c_1(2,0)}{x_1^2}t^{q-2p}\Bigr) \nonumber \\
&+& \frac{y_1^2}{x_1^2} \{1+\Bigl( 3 \frac{c_1(2,0)}{x_1^2} - 4 \frac{c_1(1,1)}{x_1 y_1} + \frac{c_1(0,2)}{y_1^2}\Bigr) t^{q-2p}\}, \nonumber \\
p y_1 t^{p-1} &\sim& \frac{ \beta }{y_1} t^{-p} \Bigl( 1 + \frac{c_1(0,2)}{y_1^2}t^{q-2p} \Bigr) \nonumber \\ 
&-& 2 \frac{y_1}{x_1} \{1+\Bigl( \frac{c_1(2,0)}{x_1^2} - \frac{c_1(1,1)}{x_1 y_1}\Bigr) t^{q-2p}\}, \nonumber
\end{eqnarray}
and for the 2-point correlation functions,
\begin{eqnarray}
q c_1(2,0) t^{q-1} &\sim& 2-2 \alpha \frac{c_1(2,0)}{x_1^2}t^{q-2p} \nonumber \\
& & -4 \frac{y_1^2}{x_1} \Bigl(\frac{c_1(2,0)}{x_1^2} -\frac{c_1(1,1)}{x_1 y_1}\Bigr) t^{q-p},\nonumber\\
q c_1(1,1) t^{q-1} &\sim& - \Bigl(\alpha \frac{y_1}{x_1} + \beta \frac{x_1}{y_1}\Bigr) \frac{c_1(1,1)}{x_1 y_1} t^{q-2p} \nonumber \\
& &+2 y_1 \Bigl(\frac{c_1(2,0)}{x_1^2}- \frac{c_1(1,1)}{x_1 y_1}\Bigr) t^{q-p}\nonumber \\ 
& &+2 \frac{y_1^3}{x_1^2} \Bigl(\frac{c_1(0,2)}{y_1^2}-\frac{c_1(1,1)}{x_1 y_1}\Bigr) t^{q-p},\nonumber\\
q c_1(0,2) t^{q-1} &\sim& 2-2 \beta \frac{c_1(0,2)}{y_1^2}t^{q-2p} \nonumber \\
& & -4 \frac{y_1^2}{x_1} \Bigl(\frac{c_1(0,2)}{y_1^2} - \frac{c_1(1,1)}{x_1 y_1}\Bigr) t^{q-p}. \nonumber
\end{eqnarray}
There are three possible pairs of exponents $(p,q)$, {\it 1}) (2/3, 1), {\it 2}) (1/2, 1), and {\it 3}) (1/2, 1/2).
The cases {\it 1}) and {\it 3}) give $x_1=y_1=0$, and the case {\it 2}) does not have a solution due to negativeness of the parameter $\alpha$.
In summary, when the parameter $\delta$ approaches to zero, the growing rates ${v_x, v_y}$ get smaller, but nonetheless $\langle x \rangle$ and $\langle y \rangle$ increase indefinitely.
On the other hand, as we have seen in the last section all the expectation values in {\it P1} tend to zero in the continuous limit $y_0 \rightarrow 0$, and we conclude the transition at the boundary {\it B1} is sudden.

Let us turn our attention to the boundary {\it B2}, where the expansion rate for $ \langle y \rangle $ vanishes as $ v_y \sim (-\lambda^B)/(1+2 v_x) $, while $ v_x \sim -\lambda $ stays finite with a negative $ \lambda $.
In this case the ratio $ r \equiv \langle y \rangle/\langle x \rangle $approaches to $0$, and coefficients of the two point functions tend to 
$$
c(2,0)\sim2,~~ c(1,1)\sim0,~~ c(0,2)\sim2 \frac{v_x}{4+v_x}.
$$
In general, coefficients of the correlation function $\langle\Delta_x^n \Delta_y^m \rangle$ behave for a small $\lambda^B$ as 
$$
c({\it even},{\it even}) \sim c({\it odd},{\it even}) \sim {\cal O}(1) ,
$$ 
and
$$ 
c({\it odd},{\it odd}) \sim c({\it even},{\it odd}) \sim {\cal O}(\lambda^B) . 
$$ 

In the case when we set $\lambda^B=0$ before taking the asymptotic limit, we expect the following type of solutions,
$$
\langle x \rangle \sim v_x t +\dots, ~~ \langle y \rangle \sim y_1 t^p +\dots,
$$ 
with $0<p<1$.
Assuming that the expansion in terms of $ \xi $ and $ \eta $ is possible, we have $\langle x \rangle\sim-\lambda t$, while from the equation for $ \langle y \rangle $, we obtain $ p=1/2$.
Since the equation for $\langle\Delta_y^2 \rangle$ gives non-vanishing $c(0,2) = 2 v_x/(4+v_x)$, the order parameter along the y-direction is $ \sqrt{\langle\eta^2 \rangle} \sim {\cal O}(t^0)$, which implies that the expansion is perturbative.
We shall consider this case in the subsequent section.
%

\section{The elongating phase ({\it P3}~)}
\setcounter{equation}{0}
\noindent
When $ \lambda<0 $, $ \lambda^B > 0$, and  $\beta > 0$, the equation for $\langle x \rangle$, eq.(4.2), gives $\langle x \rangle \rightarrow v_x t$ with $v_x = -\lambda$, assuming 
$$
\langle\frac{1}{x}\rangle \rightarrow \frac{1}{\langle x \rangle},~~ \langle\frac{y^2}{x^2}\rangle \rightarrow \frac{\langle y^2 \rangle}{\langle x \rangle^2}
$$
as $\langle x \rangle$ increases.
Then, the equation for $ \langle y \rangle $, eq.(4.3), has a stationary solution at the attractive fixed point, $ \langle y \rangle=y_1$, satisfying
$$
- \lambda^B + \beta \langle\frac{1}{y}\rangle=0 ,
$$ 
where $ y_1 $ is regarded to be a parameter at this moment.
Since $ \langle\Delta_x^2 \rangle $ also shows a linear increase in $ t $, equations of the motion for the $x$-direction are expanded in terms of the relative fluctuation $\xi$, while equations for the $y$-direction are treated without separating the fluctuation.
The set of equations of this type is named as the $\Delta$-$y$ scheme, and it is summarized in the Appendix C.

In this section we choose $ \beta $ positive, otherwise $ d\langle y \rangle/dt $ is always negative for $ \lambda^B \geq 0 $, and the motion along the y-direction becomes unstable except $\beta=-1$, which will be discussed in the next section.
Keeping only the lowest order terms of expansion along the x-direction, equations of the motion in the asymptotic region are given by
\begin{eqnarray}
\frac{d \langle\Delta_x^n y^m \rangle} {d t} &\sim&n(n-1) \langle\Delta_x^ {n-2} y^ {m}\rangle+ m (m-1+\beta) \langle\Delta_x^ {n} y^ {m-2}\rangle \nonumber \\
& &~- \frac{ 2m }{ \langle x \rangle }\langle\Delta_x^n y^m \rangle-m \lambda^B \langle\Delta_x^n y^{m-1}\rangle \nonumber \\
& &~+ [\frac{ 2m }{ \langle x \rangle^2 }\langle\Delta_x^ {n+1} y^m \rangle \nonumber \\
& &~+ \frac{n}{\langle x \rangle^2}\{\langle\Delta_x^{n-1} y^{m+2}\rangle - \langle\Delta_x^{n-1} y^m \rangle \langle y^2 \rangle\} \nonumber \\
& &~+\frac{n\alpha}{\langle x \rangle^3}\{\langle\Delta_x^{n+1} y^m \rangle-\langle\Delta_x^{n-1} y^m \rangle\langle\Delta_x^2 \rangle\}], \label{eq.6.1}
\end{eqnarray}
where the last three terms in the bracket $ [\dots]$ are reserved for the odd $ n $ case.

We start solving the $n=0$ component,
\begin{equation}
\frac{d \langle y^m \rangle} {d t} \sim m (m-1+\beta) \langle y^{m-2}\rangle -2m \frac{\langle y^m \rangle}{\langle x \rangle} -m \lambda^B \langle y^{m-1}\rangle. \label{eq.6.2}
\end{equation}
%
The solution is formally expressed as,
\begin{equation}
\langle y^m \rangle(t) \sim m t^{-2m/v_x} \int^t z^{2m/v_x} \{-\lambda^B \langle y^{m-1}\rangle(z) + (m-1+\beta) \langle y^{m-2}\rangle(z)\} dz, \label{eq.6.3}
\end{equation}
where we choose $ \langle x \rangle=v_x t $.
Carrying out the integration explicitly, the first two solutions are expressed as, 
\begin{eqnarray}
\langle y^2 \rangle &\sim& 2(1+\beta-\lambda^B y_1)\frac{t}{1+\frac{4}{v_x}} + y_2,\\
\langle y^3 \rangle &\sim& 3\{(2+\beta) y_1 - \lambda^B y_2\} \frac{t}{1+\frac{6}{v_x}} \nonumber \\
&+&\lambda^B  (1+\beta-\lambda^B y_1)\frac{t^2}{(1+\frac{3}{v_x})(1+\frac{4}{v_x})} + y_3, \label{eq.6.4}
\end{eqnarray}
where $y_2$ and $y_3$ are initial values of $\langle y^2 \rangle$ and $\langle y^3 \rangle$, respectively.
Since the expectation value $\langle y^m \rangle$ should be positive, the solution $\langle y^2 \rangle$ is acceptable only when the parameters satisfying $1+\beta-\lambda^B y_1 \geq 0$.
On the other hand, for the solution $\langle y^3 \rangle$ the positivity condition requires $1+\beta-\lambda^B y_1 \leq 0$.
Combining these conditions, $y_1$ is fixed to be $(1+\beta)/\lambda^B $.
Continuing this procedure for larger $m$, we will find that the allowed asymptotic solution, $\langle y^m \rangle \rightarrow y_m$, is required to satisfy 
\begin{equation}
(m-1+\beta) y_{m-2}- \lambda^B y_{m-1} = 0, \label{eq.6.5}
\end{equation}
which gives the asymptotic value of $\langle y^m \rangle$ to be 
\begin{equation}
y_m=\prod_{k=1}^m \Bigl(\frac{k+\beta}{\lambda^B}\Bigr). \label{eq.6.6}
\end{equation}

For an even integer $ n $ the $(n,0)$-component equation,
\begin{equation}
\frac{d \langle \Delta_x^n \rangle} {d t} \sim n (n-1) \langle\Delta_x^ {n-2}\rangle, \label{eq.6.7}
\end{equation}
gives the solution
\begin{eqnarray}
\langle \Delta_x^n \rangle &\sim& d_n t^{n/2},\nonumber \\
d_n &=& 2^{n/2} \prod_{k=1}^{n/2} (2k-1), \label{eq.6.8}
\end{eqnarray}
for $n \geq 2 $, and $ d_0=1$.
Since $y\sim {\cal O}(1)$ and $\Delta_x \sim {\cal O}(t^{1/2})$, inserting the asymptotic form, 
$$
\langle \Delta_x^n y^m \rangle=c(n,m) t^{n/2},
$$
into eq.(6.1), the leading term follows
\begin{eqnarray}
\frac{n} {2} c(n,m) t^{n/2-1}
&\sim& n (n-1) c(n-2,m) t^{n/2-1}+ m (m-1+\beta) c(n,m-2) t^{n/2}\nonumber \\
&-& \frac{ 2m }{ v_x } c(n,m) t^{n/2-1}-m \lambda^B c(n,m-1) t^{n/2}, \nonumber
\end{eqnarray}
which requires $c(n,m)$ to satisfy
\begin{equation}
0 = (m-1+\beta) c(n,m-2) - \lambda^B c(n,m-1). \label{eq.6.10}
\end{equation}
It gives the solution to be
\begin{equation}
c(n,m) = \prod_{k=1}^m(\frac{k+\beta}{\lambda^B}) c(n,0). \label{eq.6.11}
\end{equation}
From eq.'s (6.7) and (6.9) it is written as $c(n,m)=d_n y_m$, which indicates the lack of correlation between motions along $x$ and $y$ directions.

We now proceed to solve the $ n=$ {\it odd integer} case.
Since a correlation function of odd moment is not necessarily positive, we cannot utilize the positivity condition as the $n$-even case, and it is desirable to know the bound on exponents in order to select the acceptable solutions.
Making use of the Schwartz inequality, 
\begin{equation}
\langle \Delta_x^n y^m \rangle^2 \leq \langle \Delta_x^{2n} \rangle\langle y^{2m} \rangle , \label{eq.6.12}
\end{equation}
together with the previous result for $\{\langle\Delta_x^{2n}\rangle\}$ and $\{\langle y^{2m}\rangle\}$, we have the bound, 
$$
|\langle \Delta_x^n y^m \rangle| \leq {\cal O}(t^{n/2}),
$$ 
which is required asymptotically. 

Equations of the motion are given by eq.(6.1) with the additional terms reserved in $[\dots]$.
Writing the asymptotic form of solution for the even $n$ case as 
$$
\langle \Delta_x^ n y^m \rangle = d_n y_m t^{n/2},
$$ 
the contribution from additional terms is
\begin{equation}
\frac{2m}{v_x^2} d_{n+1} y_m t^{\frac{n-3}{2}}+\frac{n}{v_x^2} d_{n-1} (y_{m+2}-y_m y_2) t^{\frac{n-3}{2}}+\frac{n\alpha}{v_x^3} (d_{n+1} - d_{n-1} d_2)y_m t^{\frac{n-5}{2}}. \nonumber
\end{equation}
Since it has a different $t$ dependence in the $m=0$ case from other cases, we shall solve the equation for $m=0$ and $m\neq 0$ separately.

Firstly, we set $m=0$ in eq.(6.1), and obtain
\begin{equation}
\frac{d \langle \Delta_x^n \rangle} {d t} \sim n (n-1) \langle\Delta_x^ {n-2}\rangle +\frac{n\alpha}{\langle x \rangle^3}\{\langle\Delta_x^{n+1}\rangle-\langle\Delta_x^{n-1}\rangle\langle\Delta_x^2 \rangle\}. 
\label{eq.6.13}
\end{equation}
This equation can be solved iteratively starting from $n=3$ using the solutions for even $n$ moments, resulting
\begin{eqnarray}
\langle \Delta_x^n \rangle &\sim& \tilde{d}_n t^{(n-3)/2}\log t,\nonumber \\
\tilde{d}_n&=&\frac{2n(n-1)}{(n-3)}\tilde{d}_{n-2}, \label{eq.6.14}
\end{eqnarray}
for $n \geq 5 $, with $\tilde{d}_3=(24\alpha/v_x^3) $.
Next, we consider the $m\neq0$ cases, and we get the same relation as the previous even $n$ case:
\begin{equation}
(m-1+\beta) \langle\Delta_x^ {n} y^ {m-2}\rangle-\lambda^B \langle\Delta_x^n y^{m-1}\rangle\sim 0 .
\end{equation}
Solutions of this equation were already known to be 
\begin{eqnarray}
\langle\Delta_x^x y^m \rangle&\sim&\langle\Delta_x^n \rangle\prod_{k=1}^m \Bigl(\frac{k+\beta}{\lambda^B}\Bigr) \nonumber \\
&=&\tilde{d}_n y_mt^{(n-3)/2}\log t, \label{eq.6.15}
\end{eqnarray}
which also shows the lack of correlations between $x$ and $y$ motions.

Approaching toward the boundary {\it B2} from above, $\lambda^B \rightarrow 0^{+}$, the moments, $\langle\Delta_x^n y^m \rangle $ diverges as ${\cal O}((\lambda^B)^{-m})$.
In order to understand the nature of transition between the expanding phase and the elongating phase, we need to know properties of the boundary state along {\it B2}, which will be discussed in the subsequent section.
On the other hand $\langle\Delta_x^n y^m \rangle $ does not depend on $\lambda$ asymptotically, and the transition at the boundary {\it B3} is abrupt according to the limit in the neighboring phase {\it P1}.

\section{The polymer state and the boundary state}
\setcounter{equation}{0}
\noindent
There exist two special states for certain values of the parameter set ($\lambda$, $\lambda^B$, $\beta$).
When $ \beta $ is zero or negative, $ d\langle y \rangle/dt $ is always negative for $ \lambda^B \geq 0 $, and $ \langle y \rangle $ tends to be unphysical.
In the negative $\beta$ region there exists a special case locating at $\beta= -1$, where not only $y_2$ but also all $\{y_m\}$ are $0$ according to eq.(6.7).
In this case the distribution function $ p(x,y;t)$ approaches to 
$$
p(x,y;t)\rightarrow p(x;t) \delta (y).
$$
We may call this case as the polymer state.
In this state only the motion along the x-direction evolves as 
$ \langle x \rangle \rightarrow v_x t $ with $ v_x =-\lambda $, and 
\begin{equation} 
\langle \Delta_x^n \rangle \sim 
\left\{ 
\begin{array}{ll} d_n t^{n/2} & \mbox{if} \ n={\it even} \\ 
\tilde{d}_n t^{(n-3)/2}\log t & \mbox{if} \ n={\it odd} 
\end{array} 
\right. \label{eq.7.1}
\end{equation}
except the $n=1$ component, {\it i.e.} $\langle\Delta_x \rangle=0$.
The coefficients $\{d_n\}$ and $\{\tilde{d}_n\}$ are the same as those of the elongating phase given by eq.(6.9) and eq.(6.16), respectively.

There is another special state which appears along the boundary {\it B2}.
For an even integer $n$ the equation of motion,
\begin{eqnarray}
\frac{d \langle\Delta_x^n y^m \rangle}{d t} &\sim& n(n-1)\langle\Delta_x^ {n-2} y^m \rangle+ m(m-1+\beta) \langle\Delta_x^ n y^ {m-2}\rangle \nonumber \\
&-&\frac{2m}{\langle x \rangle} \langle\Delta_x^ n y^m \rangle, \label{eq.7.2}
\end{eqnarray}
has the solution with the product form, 
$$ 
\langle \Delta_x^n y^m \rangle \sim \langle \Delta_x^n \rangle \langle y^m \rangle,
$$ 
where each component obeys,
\begin{eqnarray}
\frac{d\langle\Delta_x^n \rangle}{d t} &\sim& n (n-1) \langle\Delta_x^ {n-2} \rangle, \\
\frac{d\langle y^m \rangle}{d t} &\sim& m(m-1+\beta) \langle y^ {m-2}\rangle -\frac{ 2m }{ \langle x \rangle }\langle y^m \rangle. \label{eq.7.3}
\end{eqnarray}
The first equation is the same as the elongating phase and we already have the solution $\langle\Delta_x^n \rangle \sim d_n t^{n/2}$ with the same $d_n$ given by eq.(6.8).
For the second equation writing $\langle y^m \rangle \sim \bar{y}_m t^{m/2}$, we obtain
\begin{equation}
\bar{y}_m = 2(m-1+\beta) \bar{y}_{m-2} - \frac{ 4 }{ v_x } \bar{y}_m. \label{eq.7.4}
\end{equation}
For an even integer $m$ we solve it iteratively with the initial value $\bar{y}_0=1$, and we get
\begin{equation}
\bar{y}_m = \Bigl(\frac{2}{1+\frac{4}{v_x}}\Bigr)^{\frac{m}{2}} \prod_{k=1}^{m/2}(2k-1+\beta),~ (m:even). \label{eq.7.5}
\end{equation}
In order to obtain solutions for an odd integer $m$ we need to give a value for $\bar{y}_1$, which is regarded as a parameter at this moment.
Then, the solution is given by 
\begin{equation}
\bar{y}_m = \Bigl(\frac{2}{1+\frac{4}{v_x}}\Bigr)^{\frac{m-1}{2}} \prod_{k=1}^{(m-1)/2}(2k+\beta)\bar{y}_1,~ (m:odd). \label{eq.7.6}
\end{equation}

For an odd integer $n$ we need to add extra-terms coming from the even $n$ solutions in the equation of motion, eq.(6.1).
Since contributions of these extra terms for $m=0$ and $m\neq 0$ have different properties similar to the previous case, we treat them separately.
For $m=0$ the equation is same as the elongating phase and the solution is known to be
\begin{equation}
\langle\Delta_x^n \rangle=\tilde{d}_n t^{(n-3)/2}\log t,
\end{equation}
with the same coefficient $\tilde{d}_n$ as eq.(6.16). 

For $m\neq 0$ case the additional terms are written by using the $n$-even solutions, $\langle\Delta_x^n y^m \rangle \sim d_n \bar{y}_m \exp\{-(n+m)/2\}$, to be 
$$
\frac{2m }{v_x^2 }d_{n+1} \bar{y}_m + \frac{n}{v_x^2}d_{n-1}(\bar{y}_{m+2} - \bar{y}_m\bar{y}_2), \label{eq.7.7}
$$
where we keep terms having the order $t^{(n+m-3)/2}$.
We note the additional contributions can be rewritten as 
$$
\frac{3+\frac{8}{v_x}}{1+\frac{4}{v_x}}\frac{m}{v_x^2}d_{n+1}\bar{y}_m,
$$
using the relationships for $d_n$, eq.(6.9). Writing the asymptotic solution as 
$$
\langle \Delta_x^n y^m \rangle\sim \hat{d}_n \hat{y}_m t^{\frac{n+m-1}{2}},
$$ 
except the $(1,0)$-component, which is $0$ by definition, the equation of motion is expressed as
\begin{eqnarray}
\frac{n+m-1}{2} \hat{d}_n \hat{y}_m &=& n (n-1) \hat{d}_{n-2} \hat{y}_m + m(m-1+\beta) \hat{d}_n \hat{y}_{m-2} \nonumber \\
&-&\frac{ 2m }{v_x} \hat{d}_n \hat{y}_m + \frac{3+\frac{8}{v_x}}{1+\frac{4}{v_x}} \frac{m}{v_x^2} d_{n+1} \bar{y}_m, \label{eq.7.8}
\end{eqnarray}
except the equation for the $(n,m)=(1,2)$ moment, which involves the $(1,0)$-component, 
\begin{equation}
\hat{d}_1 \hat{y}_2 = -\frac{ 4 }{v_x} \hat{d}_1 \hat{y}_2 + \frac{3+\frac{8}{v_x}}{1+\frac{4}{v_x}}  \frac{2}{v_x^2} d_2 \bar{y}_2. \label{eq.7.9}
\end{equation}
%
Imposing 
\begin{equation}
\hat{d}_n=2n\hat{d}_{n-2}, ~~(n: odd) 
\end{equation} 
with $\hat{d}_1=1$, eq.(7.9) becomes
\begin{equation}
\hat{y}_m = 2(m-1+\beta) \hat{y}_{m-2} -\frac{ 4 }{v_x} \hat{y}_m 
+ \frac{3+\frac{8}{v_x}}{1+\frac{4}{v_x}} \frac{4}{v_x^2} \bar{y}_m, \label{eq.7.10}
\end{equation}
where we use the relation 
$$ 
d_{n+1}/\hat{d}_n = d_{n-1}/d_{n-2} =d_2/\hat{d}_1=2 .
$$
We note eq.(7.10) becomes the same as the $m=2$ component of eq.(7.9) when we set $\hat{y}_0=0$.

The solution of eq.(7.11) is  given by
\begin{equation}
\hat{d}_n=2^{(n-1)/2} \prod_{k=1}^{(n-1)/2} (2k+1)~~(n: odd) ,
\end{equation} 
while eq.(7.12) can be solved for even $m$ and odd $m$ separately, using the solution $\bar{y}_m$ given by eq.(7.6) for the even $m$ case and eq.(7.7) for the odd $m$ case.
For an even $m$ we solve it iteratively with the condition $\hat{y}_0=0$, while for an odd $m$ the solution has the same form as the even case with the initial input value $\hat y_1$, which  we assume to be a parameter at this moment.
We give here the solutions without giving details of derivations,
\begin{equation}
\hat{y}_m = 
\left\{
  \begin{array}{ll}
 \frac{m}{2} a \Bigl(\frac{2}{1+\frac{4}{v_x}}\Bigr)^{\frac{m}{2}} \prod_{k=1}^{m/2}(2k-1+\beta) &\mbox{if} \ m={\it even}  \\
 (\frac{m-1}{2} a \hat{y}_1+\bar{y}_1) \Bigl(\frac{2}{1+\frac{4}{v_x}}\Bigr)^{\frac{m-1}{2}} \prod_{k=1}^{(m-1)/2}(2k+\beta)\bar{y}_1 &\mbox{if} \ m={\it odd},
  \end{array}
\right.
\end{equation}
where we write $a\equiv 4(3+\frac{8}{v_x})/v_x^2/(1+\frac{4}{v_x})^2$.

Until this point we obtain all quantities of our interest in the asymptotic limit with two initial input parameters, $\bar{y}_1$ and $\hat{y}_1$, which correspond to $\langle y \rangle$ and $\langle\Delta_x y \rangle$, respectively. 
Since their equations of the motion, 
\begin{eqnarray}
\frac{d\langle y \rangle}{d t} &\sim& \beta \langle y^ {-1}\rangle - 2 \frac{ \langle y \rangle }{\langle x \rangle},\\
\frac{d \langle \Delta_x y \rangle}{d t} &\sim& \beta \langle\Delta_x y^ {-1}\rangle - 2 \frac{\langle\Delta_x y \rangle}{ \langle x \rangle } 
+ \frac{\langle y^3 \rangle - \langle y \rangle \langle y^2 \rangle}{ \langle x \rangle^2 }, \label{eq.7.11}
\end{eqnarray}
involve moments with the negative exponent, it seems we are eventually forced to solve infinite towers of equations with negative integer arguments.

In order to avoid these difficulties we shall make use of the following relation for $\langle y^{-1}\rangle$,
\begin{eqnarray}
\langle \frac{1}{y} \rangle &=& \frac{ 1 }{ \langle y \rangle } \langle\frac{ 1 }{ 1+\eta }\rangle \nonumber \\
&=& \frac{ 1 }{ \langle y \rangle } \sum_{n=0}^{\infty} \sum_{m=0}^n (-1)^m \left( \begin{array}{c} n \\ m \end{array} \right) \frac{\langle y^m \rangle}{\langle y \rangle^m} \nonumber \\
&=& \frac{ 1 }{ y_1 } \sum_{n=0}^{\infty} \sum_{m=0}^n (-1)^m  \left( \begin{array}{c} n \\ m \end{array} \right) \frac{\bar{y}_m}{\bar{y}_1^m} t^{-1/2}. \label{eq.7.12}
\end{eqnarray}
Since $\bar{y}_m$ is written as eq.'s (7.6) and (7.7), eq.(7.14) indicates $\langle y^{-1}\rangle$ is expressed by $\bar{y}_1$.
From this relation and the eq.(7.12) we have the equation,
\begin{equation}
\frac{1}{2}(1+\frac{4}{v_x}) \bar{y}_1 = \frac{ \beta }{ \bar{y}_1 } \sum_{n=0}^{\infty} \sum_{m=0}^n (-1)^m  \left( \begin{array}{c} n \\ m \end{array} \right) \frac{\bar{y}_m}{\bar{y}_1^m}, \label{eq.7.13}
\end{equation}
which can, in principle, determine the parameter $\bar{y}_1$.
Similar analysis can be applied for $\langle\Delta_x y^{-1} \rangle$, giving
\begin{equation}
\frac{1}{2}(1+\frac{4}{v_x}) \hat{y}_1 = \frac{\beta }{\bar{y}_1 } \sum_{n=0}^{\infty} \sum_{m=0}^n (-1)^m  \left( \begin{array}{c} n \\ m \end{array} \right)  \frac{\hat{y}_m}{\bar{y}_1^m}+\frac{\bar{y}_3-\bar{y}_2 \bar{y}_1}{v_x^2}. \label{eq.7.14}
\end{equation}
Again, $\hat{y}_m$ is known from eq.(7.11) up to two parameters, $\bar{y}_1$ and $\hat{y}_1$, and the remained parameter $\hat{y}_1$ should be unambiguously determined. 

This concludes analyses of solutions of the master equation obtained through equations of the motion for all the possible moments, and they show the asymptotic solutions are exact and are determined without any ambiguous parameters other than $(\lambda, \lambda^B, \beta)$.  

\section{Discussion: Evolution of the universe}
\setcounter{equation}{0}
\noindent
After studying statistical properties of the 2-dimensional triangulated manifold, we are ready to introduce the space and the time as the frame of reference.
It is natural to choose space and time coordinates according to the global property of the topology.
The manifold we have constructed has the two-dimensional disk topology.
Peripheral boundary of the disk fits to the space coordinate from the symmetry of forward and backward direction.
On the other hand the direction perpendicular to the boundary is not symmetric, and we can easily recognize the inside as past and the outside as future when it is regarded as time.

Among the solutions of the equation of motion the average area, $A(\tau)$, and the average boundary length, $l(\tau)$, are the two fundamental quantities for the space and time.
Their evolution are described by the length of Markov chain, which we write $\tau$, here.
We consider these two quantities are related each other through the 'time' $t$, which we may call the physical time, as
\begin{equation}
A(t)= c \int^t l(t) dt, \label{eq.8.1}
\end{equation}
where $c$ takes into account of adjusting scales of the physical space and the physical time.

The time we defined in eq.(8.1) is a function of $\tau$, which is determined by
\begin{equation}
ct= \int^{\tau} \frac{1}{l(\tau)}\frac{dA(\tau)}{d\tau} d\tau, \label{eq.8.2}
\end{equation}
%
From the solutions, $\langle x \rangle$, and $\langle y \rangle$, we have the area and boundary length $A=\langle x \rangle/\gamma$, and $l=\langle y \rangle/\gamma$, and we can calculate the relationship between $t$ and $\tau$, for each phases as the followings.

{\it P1}: $A(\tau)\sim y_0^2/\gamma$, and $l(\tau)\sim y_0/\gamma$ for $\nu > 1$
$$
ct=0~~(\mbox{no running time}).
$$
We note that the crunch occurs as a result of collapsing.

{\it P2}: $A(\tau)=v_x \tau/\gamma$, and $l(\tau)=v_y \tau/\gamma$
$$
ct=\frac{v_x}{v_y}\log \tau, \mbox{and}~ l(t)=\frac{v_y}{\gamma} \exp\Bigl ({\frac{cv_y}{v_x}t} \Bigr )~~(\mbox{inflationary expansion}).
$$

{\it P3}: $A(\tau)=v_x \tau/\gamma$, and $l(\tau)=y_1/\gamma$
$$
ct=\frac{v_x}{y_1}\tau, \mbox{and}~ l(t)=\frac{y_1}{\gamma}~~(\mbox{stationary universe}).
$$ 

Boundary State: $A(\tau)=v_x \tau/\gamma$, and $l(\tau)=\bar{y}_1 \tau^{1/2}/\gamma$
$$
ct=\frac{v_x}{\bar{y}_1}\sqrt{\tau}, \mbox{and}~ l(t)=\frac{c\bar{y}_1^2}{\gamma v_x} t~~(\mbox{'Big Bang' universe}). 
$$
We employ the name Big Bang as the power law expansion in comparison to the inflation as the exponential expansion.

Polymer State: $A(\tau)=v_x \tau/\gamma$, and $l(\tau)=y_1/\gamma$
$$
ct=\frac{v_x}{y_1}\tau, \mbox{and}~ l(t)=0~~(\mbox{stationary universe}).
$$
The polymer universe appears when $\beta=-1$, where $\gamma^{-1}=0$

\section{Summary}
\noindent
We have started creating a two-dimensional universe under the working hypothesis, {\it There were no laws but rules at the beginning of universe}.   
The rules we have proposed consist with the three main axioms: {\it 1}) existence of the elementary unit of universe, {\it 2}) existence of neighbors of a world, and {\it 3}) equal probability of appearance of worlds. 
Based on these axioms we have obtained the master equation which describes the emergence and evolution of a two-dimensional space-time. 
The equation reveals three prominent phases, {\it i.e.} the expanding, the elongating and the collapsing phase, and two significant states, named as the polymer and the boundary state.
For each phase and state statistical averages of the area, the boundary length, and correlations between those variables and their fluctuations are obtained.
Since these solutions are expanded in terms of the order parameter, which is the inverse square root of the Markov time, they are considered to be asymptotically exact.
Moreover, solutions contain no ambiguous parameters in the asymptotic limit except three basic inputs, {\it i.e.} the cosmological constant $\lambda$, the boundary cosmological constant $\lambda^B$, and the matter content $\beta$.

The physical time is introduced from the relation between the area and the boundary length, after assigning the boundary as the physical space.
In terms of the physical time solutions of each phase show the characteristic behavior of space evolution, {\it 1}) inflationary expansion in the expanding phase, {\it 2}) stationary universe in the elongating phase, {\it 3}) point universe in the collapsing phase, {\it 4}) stringy universe in the polymer state, and {\it 5}) linear expansion in the boundary state, which we have called the Big Bang expansion.
The variety of solutions indicates the ability of describing most of the essential features of the universe even in the 2-dimensional space-time.
We remark all the possible type of time appeared in the two-dimension seems to be universal for any manifold with one boundary according to the numerical simulation for the 4-dimensional case.

In order to investigate the space-time evolution in higher dimensions we need to achieve method to enumerate triangulations.
Because of the lack of analytical means numerical efforts to obtain some hints for functional forms of the distribution function are unavoidable.
The dynamical triangulation with boundaries has been performed sometime ago for searching continuous transition \cite{Cat3, Cat4} in space-time dimension 3 and 4.
Although our purpose is to enumerate triangulated graphs of open manifolds, those investigations suggest possibikities to apply the dynamical triangulation prepared for closed spaces even for open cases.

To accomplish the creation of universe it is the most important to answer following two questions among many.
Firstly, how the three parameters are determined at the very beginning?
Part of this question will be solved by moving up to higher dimensions.
In a four-dimensional universe, where the conformal symmetry is broken and the cosmological constants are not constant any more, the effective cosmological constants will be allowed to make time evolution.
In the 2-dimension the cosmological constants obtained by the matrix model are large, and the universe is in the expanding phase. 
If the conformal symmetry is broken, these two parameters are not constant anymore, and they will converge to the small fixed point in the parameter space. 
For the matter part of parameters $\beta$ we may say that any combination of fields is possible.
In case we are asked to explain the raison df?tre of the matter content of this universe, we need to accept the anthropological principle at this moment.

Second problem is the birth of quantum dynamics, which we think the primordial dynamics of universe.
In contrast to the master equation which is the same type as the diffusion equation, dynamical equation is non-diffusive.
It is the hardest part of problem to answer how the complex function becomes inevitable in the course of evolution of space and time.
Unfortunately, we do not have any answer to this question.

In spite of the lack of quantum dynamics we consider the creation of space-time can stand by itself.
Same kind of situation arises in evolution of metals: firstly a metal lattice is formed which is dominantly the classical statistical process. Later, conduction of electrons and the electron-phonon interaction emerges in the metal which are described by quantum mechanics.
Afterall, if the evolution of universe at its beginning is inflational, the quantum dynamical correlation cannot influence the space-time dynamics beyond its limited event horizon, and the evolution of space becomes inevitably statistical.

\begin{center}
   {\Large {\bf Acknowledgments}}
\end{center}
I am in debt to professor A. Iso and K. Hamada of the theory center(KEK) for their encouragements. 
I thank Dr. Horata for his assists to complete the manuscript.

\appendix
\vspace{5mm}
\begin{center}
   {\Large {\bf Appendix}}
\end{center}
\section{The master equation in the asymptotic limit}

\setcounter{equation}{0}
\noindent
The Markov process in the simplicial space is described in terms of the descrete variables ($ N_2$,$\tilde{N_1}$) and the descrete time $n$:
\begin{equation} 
\tilde{p}_a [n+1] - \tilde{p}_a [n] =  \sum_{b(a)} \{\tilde{p}_b [n] \tilde{w}_{b \rightarrow a}-\tilde{p}_a [n] \tilde{w}_{a \rightarrow b} \}, \label{eq.A.1}
\end{equation}
where $\tilde{w}_{a \rightarrow b} = \{ 1 + exp(S_b - S_a)\}^{-1}$ with
\begin{equation}
S_a = S(N_2,\tilde{N_1}) = - \alpha \log N_2 - \beta \log \tilde{N}_1 + \gamma \frac{\tilde{N}_1^2}{N_2}  + \mu N_2 + \mu^B \tilde{N}_1. \label{eq.A.2}
\end{equation} 
Here, the sum over $b(a)$ is restricted within the four neighbor points of $a=(N_2,\tilde{N}_1)$ connected by the four moves, ${\cal M}_1:(N_2 + 1, \tilde{N}_1 +1)$, ${\cal M}_2:(N_2 - 1, \tilde{N}_1 +1)$, ${\cal M}_3:(N_2 + 1, \tilde{N}_1 -1)$, ${\cal M}_4:(N_2 - 1, \tilde{N}_1 -1)$.
Introducing the three scale parameters $(A_0, l_0, \tau_0)$ for the physical variables, the area $ A = A_0 N_2$, the boundary length $l=l_0 \tilde{N}_1$, and the Markov time $ \tau = \tau_0 n$,
the probability distribution function is translated as
$$ \tilde{p}_{(N_2,\tilde{N}_1)}[n] \sim p(A,l;\tau). $$
In the continuum limit, where $(A_0, l_0, \tau_0) \rightarrow 0$, the finite shift of a veriable of a function, like $ f(x+x_0)=exp(x_0  d/dx) f(x)$, is expanded as $ \{ 1+x_0 d/dx+(1/2)(x_0 d/dx)^2+\cdots \} f(x)$.
In the l.h.s of eq.(A.1) the lowest order surviving term is ${\cal O}(\tau_0)$, while the r.h.s of eq.(A.1) the expansion starts with the second order in $A_0$ and $l_0$.
We shall keep only those leading terms for the continuum limit assuming ${\cal O}(\tau_0) \sim {\cal O}(A_0^2) \sim {\cal O}(l_0^2)$ to obtain
\begin{equation}
\tau_0 \frac{\partial}{\partial \tau} p[A,l;\tau] \\
=\Bigl\{A_0^2 \Bigl(\frac{\partial^2}{\partial A^2} + \frac{\partial S_a}{\partial A} \frac{\partial}{\partial A} + \frac{\partial^2 S_a}{\partial A^2}\Bigr) + l_0^2 \Bigl( \frac{\partial^2}{\partial l^2}+  \frac{\partial S_a}{\partial l} \frac{\partial}{\partial l}+\frac{\partial^2 S_a}{\partial l^2}\Bigr ) \Bigr\} p[A,l;\tau]. 
\label{eq.A.4}
\end{equation}
It is free to choose the scale parameters so that they satisfy $\tau_0=A_0^2=l_0^2$.

\section{The $\Delta$-$\Delta$ scheme}

\setcounter{equation}{0}
\noindent
When the two parameters $ \lambda $ and $ \lambda^B $ are chosen in the region of the expanding phase, which locates mostly in the third quadrant and partially in the fourth quadrant, the area and the boundary length increase at their maximum rates in time. 
Variables $x$ and $y$ are then separated to mean values $ \langle x \rangle $, $ \langle y \rangle $ and fluctuations $\Delta_x=x-\langle x \rangle$ and $\Delta_y=y-\langle y \rangle$.
Since the second moments of fluctuations $\langle\Delta_x^2 \rangle$ and $\langle\Delta_y^2 \rangle$ also increase proportional to time, the relative fluctuations $\xi \equiv \Delta_x/\langle x \rangle$ and $\eta \equiv \Delta_y/\langle y \rangle$ are of the order {\cal O}($t^{-1/2}$).
The equation of motion for averages of $x$, $y$ are written as
\begin{eqnarray}
\frac{d \langle x \rangle} {d t} &=& - \lambda + \frac{\langle y \rangle^2}{\langle x \rangle^2}\langle\frac{(1+\eta)^2}{(1+\xi)^2}\rangle + \frac{\alpha }{\langle x \rangle}\langle\frac{1}{1+\xi}\rangle ,\\
\frac{d \langle y \rangle} {d t} &=&  - \lambda^B - 2 \frac{\langle y \rangle}{\langle x \rangle}\langle\frac{1+\eta}{1+\xi}\rangle  + \frac{\beta}{\langle y \rangle}\langle\frac{1}{1+\eta}\rangle, \label{eq.B.2}
\end{eqnarray}
and moments of fluctuations $\langle\Delta_x^n \Delta_y^m \rangle$ as
\begin{eqnarray}
\frac{d \langle\Delta_x^n \Delta_y^m \rangle}{d t} &=& n(n-1) \langle\Delta_x^ {n-2} \Delta_y^ {m}\rangle+ m(m-1) \langle\Delta_x^ {n} \Delta_y^ {m-2}\rangle \nonumber \\
&+&n \frac{\alpha}{\langle x \rangle} \langle \Delta_x^{n-1} \Delta_y^{m} \Bigl\{ \frac{1}{1+\xi} - \langle\frac{1}{1+\xi}\rangle \Bigr\}\rangle \nonumber \\
&+&n \frac{\langle y \rangle^2}{\langle x \rangle^2} \langle \Delta_x^{n-1} \Delta_y^{m} \Bigl\{ \frac{(1+\eta)^2}{(1+\xi)^2} - \langle\frac{(1+\eta)^2}{(1+\xi)^2}\rangle \Bigr \}\rangle  \nonumber \\
&+&m \frac{\beta}{\langle y \rangle} \langle \Delta_x^{n} \Delta_y^{m-1} \Bigl\{ \frac{1}{1+\eta} - \langle\frac{1}{1+\eta}\rangle \Bigr\} \rangle \nonumber \\
&-&2m \frac{\langle y \rangle}{\langle x \rangle} \langle \Delta_x^{n} \Delta_y^{m-1} \Bigl\{ \frac{1+\eta}{1+\xi} - \langle\frac{1+\eta}{1+\xi}\rangle \Bigr \} \rangle. \label{eq.B.3}
\end{eqnarray}

\section{The $\Delta$-$y$ scheme}

\setcounter{equation}{0}
\noindent
When the boundary cosmological constant changes its sign ({\it i.e.} $ \lambda^B  > 0 $ ), the boundary length stops increasing and the universe elongates as time goes by.
In this case expanding the equation for fluctuations along the $ y-$direction cannot be justified, while the motion along $x$-direction keeps expanding with its maximum rate.
Then, we shall consider equations of the motion without splitting $<y>$ and $<\Delta_y$, and we obtain the equation for the average $\langle x \rangle$, 
%
\begin{equation}
\frac{d \langle x \rangle} {d t} = - \lambda  + \frac{1}{\langle x \rangle^2}\langle\frac{y^2}{(1+\xi)^2}\rangle + \frac{\alpha }{\langle x \rangle}\langle\frac{1}{1+\xi}\rangle, \label{eq.C.1}
\end{equation}
and the correlation $\langle \Delta_x^n y^m \rangle $,
\begin{eqnarray}
\frac{d \langle\Delta_x^n y^m \rangle}{d t} &=& n(n-1) \langle\Delta_x^ {n-2} y^ {m}\rangle+ m(m-1+\beta) \langle\Delta_x^ {n} y^ {m-2}\rangle \nonumber \\
&+&n \frac{\alpha}{\langle x \rangle} \langle \Delta_x^{n-1} y^{m} \Bigl\{ \frac{1}{1+\xi} - \langle\frac{1}{1+\xi}\rangle \Bigr\}\rangle \nonumber \\
&+&n \frac{1}{\langle x \rangle^2} \langle \Delta_x^{n-1} y^{m} \Bigl \{ \frac{y^2}{(1+\xi)^2} - \langle\frac{y^2}{(1+\xi)^2}\rangle \Bigr \}\rangle \nonumber \\  
&-&2m \frac{1}{\langle x \rangle} \langle \Delta_x^{n} y^{m} \frac{1}{1+\xi} \rangle - m \lambda^B \langle \Delta_x^n y^{m-1}\rangle. \label{eq.C.2}
\end{eqnarray}


\end{document}